# Light emission from particle beam induced plasma – An overview

A. Ulrich

Physik Department E12, Technische Universität München, James-Franck-Str. 1, 85748 Garching, Germany

**Abstract.** Experiments to study the light emission from plasma produced by particle beams are presented. Fundamental aspects in comparison with discharge plasma formation are discussed. It is shown that the formation of excimer molecules is an important process. This paper summarizes various studies of particle beam induced light emission and presents first results of a direct comparison of light emission induced by electron- and ion beam excitation. Both high energy heavy ion beam and low energy electron beam experiments are described and an overview over applications in the form of light sources, lasers, and ionization devices is given.

## 1. Introduction

In contrast to most places in the universe a plasma here on Earth has to be formed from neutral atoms or molecules. Nature shows us two ways to form a plasma in the Earth's atmosphere: the various forms of lightning and the aurora phenomena in the upper atmosphere around the magnetic poles. Counterparts of these two plasmas in the laboratory are the various forms of discharges and particle beam induced plasmas, respectively. Aspects of particle beam induced plasmas are discussed in this paper. It will focus on light emission since this allows a deep insight into the excitation and de-excitation processes and is also a guideline for practical applications in the form of coherent and incoherent light sources. The concept of a long term research program is described. Earlier work is cited and selected topics of ongoing research are discussed in more detail.

Particle beam induced light emission is a phenomenon which is probably more frequently encountered and used than one might think at first glance. All the cathode ray TV tubes, for example, make use of electron beam induced scintillation of matter. Energetic particles emitted from radioactive material had been used for a long time to illuminate the scales of clocks and other instruments in the dark. So called "extended air showers" induced in the atmosphere by high energy cosmic rays are accompanied by an invisible scintillation in the ultraviolet spectral range (Kegl et al., 2008). A wide variety of solid and liquid material such as NaI, CaWO$_4$, and various liquid scintillators is used for particle detection and identification. Scintillation detectors, however, mostly work with light pulses produced by individual particles, not particle beams. But particle beams can be helpful in this field to specify the emission spectrum since they produce enough scintillation light that wavelength resolved studies can be performed with reasonable signal to noise ratios (Undagoitia et al., 2009; Undagoitia et al., 2010). A not so well known field of particle beam induced plasmas can be described as nuclear induced plasmas and nuclear pumped lasers in which a high flux of fission fragments is used to pump gas lasers (Shaban et al., 1993). Early work performed by the author was based on this work and lead to the first heavy ion beam pumped laser in which a beam of 100 MeV sulfur ions was used to p2ump a laser (Ulrich et al., 1983). More recent work of this type has demonstrated that this concept could be a route to short wavelength lasers (Ulrich et al., 2006). Motivated by ion beam experiments, the author and his coworkers have paved the way to a new field of particle beam induced light emission in which low energy electrons are used to excite and ionize gases and liquids (Wieser et al., 1997). Some of the applications of this enabling technology will be described below.

General aspects of particle beam induced plasmas are described in the next paragraph. Experimental techniques will be discussed in paragraph 3. Then specific experiments with high energy heavy ion- and low energy electron beams will be presented in paragraph 4 and 5, respectively. An outlook and possible developments in the field will finally be given in section 6.

## 2. Fundamental aspects of particle beam induced plasmas

There is a great variety of ways to produce a plasma from flames to the interaction of extremely intense laser fields 2with matter. Discharges, however, are the most frequently used way to form a plasma for experiments in the laboratory, in lighting devices, and industrial processes like plasma etching. However, there is a fundamental difference between a discharge-produced plasma and a particle beam induced plasma! A plasma is characterized by its degree of ionization represented by the density of free electrons and the energy distribution function of these electrons. The latter describes the probability of an electron to have a certain kinetic energy. In a discharge the electrons are accelerated by an electric field, constantly gaining kinetic energy in a dc discharge or at least during a certain phase of an ac- or pulsed discharge. In a particle beam induced plasma, in contrast, the electric field by which charged particles gain their kinetic energy is geometrically decoupled from the plasma volume. This avoids what the author calls the "Franck Hertz barrier". As is well known from the Franck Hertz experiment, electrons gain energy in a discharge only until they have acquired enough kinetic energy to induce an inelastic collision in which they lose again most of their kinetic energy. Therefore, only the "high energy tail" of the electron energy distribution function will be able to ionize the material in a single step in which the electrons are accelerated. The entanglement of energy gain and loss by the free electrons with their mean free path and the material- and energy dependent collision cross sections for excitation and ionization leads to the problems of discharge ignition and the run- away effects when the discharge has ignited.

All these problems disappear when the acceleration of the particles which excite and ionize the medium is performed outside the medium. The plasma forming particles as well as all secondary charged particles formed in collision processes start out with high energy and will only lose energy until they are stopped in the medium or exit the medium. This has the great advantage, in particular for systematic scientific studies, that the density and composition of the medium can be freely chosen. Gas targets, for example, can be used from almost zero pressure up to way above atmospheric pressure without any fundamental changes in the excitation scheme. The primary projectiles will always be stopped by the same amount of target atoms or molecules. No arc or streamer will form at elevated pressure leading to inhomogeneous excitation with mechanisms which are fundamentally different from a glow discharge, and the absence of electrodes avoids contamination of the plasma by erosion of the electrode material.

Many experiments and applications with particle beam excitation use rare gases as the target material for the obvious reason that they are easy to be handled, experimentally, and that permanent changes of the target material by chemical reactions are avoided. In this context particle beam excitation has a specific advantage: High lying energy levels of target species can be collisionally excited by the high energy projectiles. The energetically high lying excited levels and the high ionization energies of the rare gases ranging from 12.1 to 24.6 eV from xenon to helium, in particular, can be reached easily in a cold gas. Particle beams also lead to a homogeneous excitation avoiding filamentation which occurs in discharges at high gas densities. This is the background why particle beam excitation of rare gases is ideal for a field of research called excimer physics (Rhodes, 1984). "Excimers" (excited dimers) are molecules formed from atoms which have a repulsive potential in the ground state but can form bound molecules in electronically excited states. Rare gases show this behavior. The possibility to populate high lying levels by particle beam excitation in a low temperature gas avoids thermal dissociation of the excimer molecules.

For some fundamental studies and for many applications it is also important to obtain a high power density in the medium. For particle beam excitation with a given beam current this is achieved when the energy transfer from the projectile to the target material per unit length along the projectile's trajectory is large. This specific energy loss dE/dx of the projectile is described for charged particles by the Bethe formula (Bethe, 1930).

The dominating term in this formula scales with $z^2/v^2$, where z is the effective charge of the projectile and v its velocity. This aspect is the background why both high energy heavy ions with high effective z and low energy electrons with z = 1e but a low velocity v are used for the experiments and applications described below. Another aspect which is related to the energy loss issue of the projectiles is the technique by which the projectiles enter, for example, a gas target. The projectiles are sent through an entrance foil into the gas (or a liquid) for that purpose and they should deposit only a small fraction of their kinetic energy in that foil. For projectile velocities which correspond to a region beyond the so called Bragg- peak the $1/v^2$ scaling is very helpful since the energy loss is low for the high velocity of the projectiles traversing the foil and then increases in the gas target with the slowing down of the projectiles. This means that the Bragg- peak lies within the target gas. Practical aspects of the particle beam plasma experiments are described in the next paragraph.

## 3. Experimental techniques

Since the technology for accelerating both ion- and electron beams is well established, the primary experimental issue here is to send ion- or electron beams into gases or liquids. Since the energy loss of the projectiles scales with the atomic number Z of the target material, low Z material is preferentially used for the entrance foil. A technique in which titanium foils are rolled down to a thickness of 2.44 μm (1.2 mg/cm$^2$) and glued onto a flange with typically 4 mm diameter open aperture has been developed for heavy ion beam experiments with projectiles of typically 3 MeV/u kinetic energy (Ulrich et al., 1983; Ribitzki et al., 1994). Much thicker entrance foils can be used for heavy ion experiments performed at the synchrotron SIS at the Helmholtzzentrum für Schwerionenforschung (GSI) near Darmstadt, Germany. There, uranium projectiles with 300 MeV/u kinetic energy are routinely used for plasma physics experiments (Hoffmann et al., 2005; Ulrich et al., 2006).

Much more delicate entrance foils have to be used for low energy (~12 keV) electron beams. The author and his coworkers have developed a technique in which very thin ceramic membranes are used for that purpose. Silicon nitride and combinations of silicon nitride and silicon oxide of 300 nm total thickness has so far been successfully used for that purpose. The energy loss in the entrance foil of 12 keV electrons, an energy frequently used for the devices, is on the order of 15%. Details are described in ref. (Morozov et al., 2008b). These thin membranes produced by solid state technology on a silicon wafer have several advantages. The reduction of the electron energy from 100 keV and more, which is the range where traditional electron beam devices such as electron beam pumped lasers work, down to ~10 keV leads to a strong increase of power deposition density for a given beam current as described above. This is due to the very small volume in which the beam power is deposited. In the case of 1 bar argon or air all the energy in a ~10 keV beam exiting the entrance foil is coupled into a volume of only about 1 mm$^3$. Beam pulses of 2 A current over 50 ns were already demonstrated which corresponds to a power density on the order of 20 MW/cm$^3$ deposited in the target gas at 1 bar. This is a similar value as provided by the most intense heavy ion beam pulses presently available, however, of course in a much smaller volume. The low particle energy has the advantage that no hard x-rays are produced so that easy to use, compact, portable devices which make use of electron beam excitation of gases and liquids can be designed for the first time by the enabling technology of thin ceramic entrance foils.

Time resolved optical spectroscopy is used to study the particle beam induced light emission. This is a standard technique. Reflective imaging optics is used to collect light emitted from the target. The light is analyzed with grating monochromators. Wavelength spectra are recorded by scanning the monochromator. Time spectra are recorded by using pulsed excitation, setting the monochromator to a specific wavelength and measuring the time dependence of the light intensity. Phototubes are normally used as the detector. A gated image intensifier with a diode array and small grating spectrometers with fiber optics coupling are also used for optical spectroscopy, where applicable. The wavelengths which are covered in our experiments range from about 60 nm to 1000 nm. An extension into the x-ray region and the near infrared is in preparation. More information about the techniques can be found in ref. (Morozov et al., 2005a; Morozov et al., 2006; Ulrich et al., 2009).

## 4. Heavy ion beam induced plasmas and their application

Two types of heavy ion beams are used for studying the ion beam induced light emission: The Munich 15 MV Tandem van de Graaff accelerator providing ions from hydrogen to uranium with typically 3 MeV/u particle energy for medium heavy ions and the heavy ion synchrotron SIS at the Helmholtzzentrum für Schwerionenforschung (GSI), Darmstadt providing e.g. $^{238}$U beams with 300 MeV/u. The Tandem Accelerator can be operated fully dc as well as in pulsed modes with a minimum pulse length of 2 ns. The uranium beam at GSI is pulsed either in a mode with 4 equally spaced bunches in about 1 μs or with individual pulses of 100 ns duration. The pulse-train of 4 bunches or the short pulses are normally launched as "single shot" events. Very high power levels can be achieved in this mode by more than $2 \times 10^9$ ions compressed into such a single pulse. This corresponds e.g. to ~23 J pulse energy, ~230 MW beam power and with a focus spot size smaller than 1 mm diameter more than 12 MW/cm$^3$ deposited in 1 bar argon gas for the uranium ions mentioned above.

The high absolute power and specific energy loss leading to a high power density delivered to an elongated target region by heavy ion beams makes this excitation technique attractive for pumping gas lasers. Heavy ion beam pumped laser operation has been demonstrated with both types of beams in earlier experiments. These laser experiments are well documented in the literature (Ulrich et al., 1983; Ulrich et al., 1988; Ulrich et al., 1990; Ulrich et al., 1993; Ulrich et al., 1994; Ulrich et al., 2006). The ~100 MeV sulfur beam has also been used for various gas kinetic studies (Ribitzki et al., 1994; Salvermoser et al., 1998). Here I will describe a recent experiment in which the technique of ion beam and electron beam excitation has been combined for the first time. The intention of these experiments is to study the mechanism of heavy ion beam induced light emission in more detail by disentangling the light production via the primary heavy ion projectiles and the secondary electrons produced in the slowing down process of the ions. It had been realized that the particle energy of the secondary electrons produced by a ~3 MeV/u $^{32}$S ion beam cover an energy range up to several keV (Ribitzki et al., 1994). The technique of sending 12 keV electrons into gas targets can therefore be used to simulate these secondary electrons. The concept is to overlap an ion beam and an electron beam in the identical target gas at the same spot. With this setup it is possible to, so to say, add extra "secondary" electrons to the heavy ion beam induced plasma and to study the effects of this combination. First results from an experiment in which two dc beams are combined are described here.

Rare gases were used for the experiments. Most of the data presented here were obtained with argon as the target gas. An overview spectrum of light emitted from argon at 600 mbar irradiated by a beam of 120 MeV $^{32}$S ions is shown in Fig. 1. The wavelength range of the spectrum is limited by the cutoff wavelength of optical windows near 110 nm at short wavelengths and by the wavelength limit in sensitivity of the phototube used as the detector at the long wavelength side of the spectrum. Most of the emission features can be explained rather easily. The intense peak at ~130 nm is the so called second continuum of the excimer emission (Rhodes, 1984). The so called "classical left turning point" (LTP) is an emission from high lying vibrational levels of neutral excimer molecules (Krötz et al., 1991; Krötz et al., 1993). Line radiation from atoms and ions and a weak emission from a H$_2$O impurity leading to OH$^*$ emission (Morozov et al., 2005b) can be found at longer wavelengths. The so called "third excimer continuum" extends from approximately 170 to 400 nm. It is related to the radiative decay of ionic excimers such as Ar$_2^{2+}$ which form from Ar$^{2+}$ atomic ions (Wieser et al., 2000). Since the cross sections for producing Ar$^{2+}$ ions with heavy projectiles is larger than with electrons it can be expected that the intensity ratio of third continuum radiation and second continuum radiation (neutral excimer molecules) is higher for ion beam excitation compared with electron beam excitation. This effect has actually been observed. It is shown in Fig. 2. A comparison of spectra emitted from 300 mbar argon gas excited by a dc 120 MeV $^{32}$S beam and a dc 12 keV electron beam, respectively, is plotted in this figure. The two spectra were recorded using the identical setup and the identical target gas right after each other. They are normalized at the second continuum. The higher intensity of the ionic feature (third continuum) with ion beam excitation is obvious. Similar effects are expected for the comparison of line intensity for neutral and ionic lines. In the data available to date the effect is surprisingly small or completely absent for ArII and ArI lines. This will have to be studied in detail in forthcoming experiments.

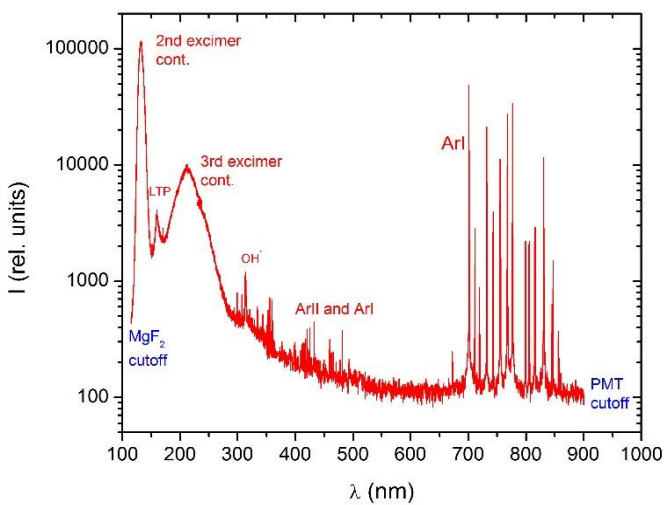

*Fig. 1.* This spectrum provides an overview over the light emitted from argon gas at a pressure of 600 mbar over a wide spectral range from ~100 to ~1000 nm. The gas was excited by a dc 120 MeV $^{32}$S beam. The energy of the projectiles at the place of light emission is ~100MeV. In the short wavelength range the spectrum is dominated by structures originating from the decay of excimer molecules, the second continuum, the "classical left turning point" (LTP), and the "third continuum". The OH* emission marks a small remaining water impurity (Morozov et al., 2005b). Spectral lines of argon atoms and ions appear on the long wavelength side.

There is also a practical application of the light emitted from ion beam excited target gas. The background is that conventional beam profile measurements fail for the very intense ion beams which have become available. Solid state devices such as metal grids and scintillator material are destroyed by the high power and energy deposition of these beams. Gas targets have the obvious advantage that they are not destroyed by the beam and the light emission can be used to locate the beam. There is, however, not a one to one correlation between beam profile and light emitting volume. The issues to be considered are the secondary electrons which excite target material outside the beam, diffusion of excited atoms or molecules and shock waves modifying the density of the gas around the beam axis. These issues are discussed in more detail in ref. (Varentsov et al., 2008) and will be described for recent experiments in a forthcoming publication.

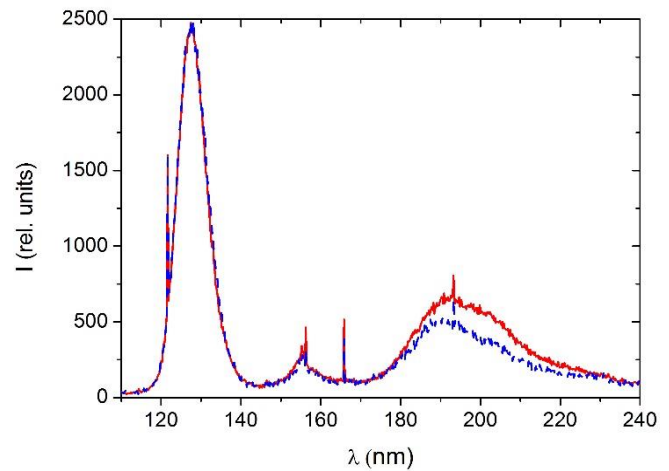

*Fig. 2.* Two spectra emitted from argon at 300 mbar are shown. The spectrum shown as a dashed (blue) line was recorded with 12 keV electron beam excitation and the spectrum shown with a solid (red) line was recorded with 120 MeV $^{32}$S beam excitation. The difference in intensity in the so called "third excimer continuum" around 200 nm, originating from ionized target species, is attributed to direct excitation by the sulfur projectiles which have a higher cross section for forming multiply ionized target species than electrons.

## 5. Low energy electron beam induced plasmas and their application

Although the low energy electron beam technique had originally been developed for the fundamental studies described in paragraph 4 it was quickly realized that they are also very useful for other applications. First of all, electron beam excitation of dense rare gases can be used as a brilliant light source in the vacuum ultraviolet (VUV) spectral region due to the favorable excitation conditions for excimer molecules described in paragraph 2 (see also the peak near 130 nm in Fig. 1). Efficiencies for the conversion of electron beam power into VUV light of 31%, 32%, 42% and 42% have been measured for the so called second continuum radiation at 83, 126, 150, and 172 nm in Ne, Ar, Kr, and Xe, respectively (Morozov et al., 2008a). An overview over the technology can for example be found in ref. (Ulrich et al., 2009) and references therein. Here I will describe a recent improvement: a brilliant, intense VUV light source covering a previously almost inaccessible spectral range for portable VUV sources (Dandl, 2010, Dandl, 2011). It is based on a table top electron beam sustained discharge.

The concept of electron beam sustained discharges is the combination of electron beam and discharge excitation of gases for plasma formation. This approach has previously for example been used to pump molecular gas lasers (Reilly 1975). Here we have combined the technology to couple low energy electron beams into gases, as described above, with an RF (radio frequency) discharge. Argon gas of typically 1.3 bar gas pressure was irradiated by a 12 keV, 5 µA electron beam. An aluminum electrode was placed at a distance of ~3 mm in the gas in front of the ceramic membrane. An RF generator capable of producing up to 30 W RF power at a frequency of 2.45 GHz was connected to the electrode via an RF cable and a directional coupler for measuring the RF power which was actually coupled into the plasma. It was found that the electron beam sustained discharge develops into a self maintained discharge with increasing RF power, emitting vacuum ultraviolet light with characteristics which are very promising for applications, both with respect to spectral shape and absolute intensity.

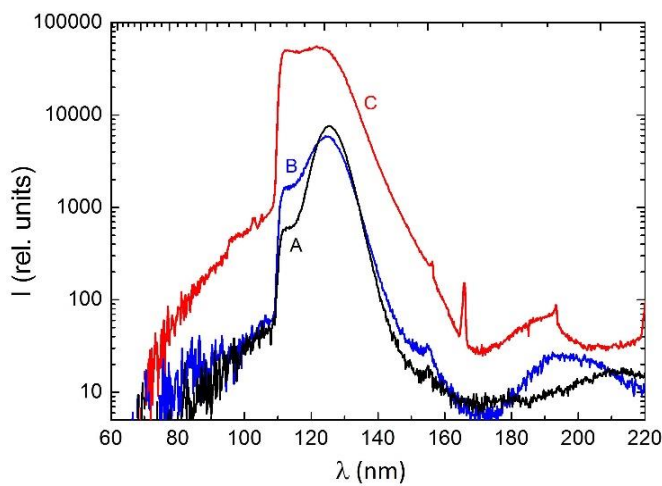

*Fig. 3:* *The three spectra shown correspond to pure electron beam excitation (A), electron beam excitation with RF power added (B), and 6 W RF excitation (C). Note the changes in the spectral shape of the emission and the logarithmic intensity scale. The RF induced spectrum matches very well with the cutoff wavelength of LiF and MgF$_2$ thereby representing a light source emitting with high intensity and efficiency at the shortest wavelengths possible for a setup containing optical windows.*

The main effect induced by the additional power deposited in the pre- ionized gas containing excimer molecules is a redistribution of the population density in the vibrational levels of the excimer molecules. It is assumed that the RF field accelerates free electrons so that the electron energy distribution function in the target gas is modified in such a way that the number of electrons with ~1 eV kinetic energy is enhanced and that collisions of these electrons with the excimer molecules populate the higher vibrational levels. Heating of the gas by the RF power and related collisions with atoms in the target gas will act in the same direction. The population of higher lying vibrational levels leads to optical transitions from these levels, which in turn leads to a significant modification of the emission spectrum. Since the high lying vibrational levels are energetically close to the energy of the resonance line, the spectrum is filled with intensity between the resonance line and the second excimer continuum. This is demonstrated for the case of argon in Fig. 3. Details of the physics and technology of table top electron beam sustained and electron beam ignited RF discharges are documented in a thesis by T. Dandl (Dandl, 2010) and will be presented in forthcoming publications (Dandl, 2011).

An important application of brilliant VUV (vacuum ultraviolet) light sources is photoionization. The so called "soft" photoionization (SPI) of analyte molecules in mass spectrometers is of particular interest. This has been demonstrated and discussed e.g. in the following publications (Mühlberger et al., 2002; Mühlberger et al., 2005a; Mühlberger et al., 2005b). With an energy on the order of 10 eV, the VUV photons can ionize most of the more complex molecules in a single step by so called "single photon absorption ionization" (SPI). The advantage of SPI is that the analyte molecules are normally not destroyed by the ionization process which is a problem for ionization by (70 eV) electron collisions which is a standard technique in most mass spectrometers. Another advantage of SPI is a partial selectivity of the detection process since only molecules with ionization energy lower than the photon energy are ionized and detected. Nitrogen and oxygen molecules in air as a carrier gas e.g. are not yet ionized by ~10 eV photons. In summary SPI leads to very clean mass spectra also for complex compositions of analyte molecules such as car exhaust, fumes emitted by coffee roasting, human breath etc. Direct ionization of gas at atmospheric pressure in so called "ion mobility spectrometers" (IMS) is another application of the low energy electron beam technique (Gunzer et al., 2010).

Besides applications for fundamental research such as gas kinetic studies (Morozov et al., 2005a) the technique of low energy electron beam excitation can help to characterize scintillation detectors which are widely used for particle detection. Applications in this field have so far concentrated on particle detection in particle astro physics (Morozov et al., 2008c; Pereira et al., 2010). The advantage of using a low energy electron beam is that the experiments can be performed with a table top setup with actual particle excitation but without shielding problems associated e.g. with tests using radioactive sources. The relatively high beam currents are of particular advantage for studying the spectral shape of the scintillation light. This has been demonstrated for the scintillation of air which is used for detecting so called extended air showers induced by high energy cosmic rays (Morozov et al., 2008c; Pereira et al., 2010) and scintillator material planned to be used for dark matter detection (Undagoitia et al., 2009; Undagoitia et al., 2010). Liquid rare gases used for particle detection, in particular, are closely related to the light sources described above. They are presently being studied with electron beam excitation in this context (Heindl et al., 2010; Heindl et al., 2011). The high specific power deposition of low energy electrons has also been used to demonstrate a smaller than table top transversely pumped electron beam pumped laser (Skrobol et al., 2009).

## 6. Summary and Outlook

There is obviously a great potential for future fundamental plasma physics studies as well as practical applications using particle beam induced plasmas and their light emission in various spectral ranges. The very high power densities which can be achieved with the heavy ion beams at GSI have the potential to study the transition from cold to hot plasmas in great detail. Saturation effects in producing excimer molecules and short wavelength laser schemes can be studied. Understanding the light emission processes in detail will also be helpful to determine the actual beam diameters in the focal spot of high intensity beams. The experiments with low power ion beams in combination with electron beam excitation are only at their beginning. Using pulsed beams and various time delays between the pulses should provide profound information about the excitation and de-excitation processes in the target.

A major technological step is planned for the low energy electron beam technique. This is to replace the hot cathode electron guns by electron sources using field emission for producing the electron beam. Carbon nanotube (CNT) field emitter systems will be tested for that purpose. This will avoid the extra power which is needed for heating the cathode and will thereby allow very compact devices with low power consumption e.g. for hand held systems to be developed. If such devices can be mass produced one can think of applications for driving chemical reactions with electron beams. The potential advantages of such an approach are discussed in ref. (Ulrich et al., 2010) which shows that low energy electron beam excitation may become an additional tool for performing plasma- chemistry.

## Acknowledgement

This overview over a long term project would not have been possible without all the coauthors of the papers of our group which have been cited. The author is also thankful for the help of engineers, technicians and accelerator operators who contributed to this work. All funding agencies which have supported the projects are also gratefully acknowledged, long term funding by the Maier-Leibnitz-Laboratory for Nuclear-, Particle- and Accelerator-Physics of the Ludwig-Maximilians-Universität München and the Technischen Universität München (MLL) and the German Bundesministerium für Bildung und Forschung BMBF in particular. BMBF funding is presently provided by the projects No. 13N9528 and 13N11376. The author thanks Thomas Heindl for his help in preparing the manuscript.